\documentclass[aps,physrev,twocolumn,longbibliography,nobalancelastpage]{revtex4-2}
\usepackage{amsmath}
\usepackage{graphicx}
\usepackage{physics}
\usepackage{comment}
\usepackage{hyperref}
\usepackage{braket}
\usepackage{soul,color}
\usepackage{bbold}
\usepackage{braket}

\usepackage{xcolor} 
\definecolor{mypink1}{rgb}{0.858, 0.188, 0.478}
\definecolor{titlecolor}{RGB}{74, 114, 159}
\definecolor{titledarkcolor}{RGB}{51,102,153}
\definecolor{LightGrey}{RGB}{232, 232, 232}
\definecolor{Grey}{RGB}{222, 223, 225}
\definecolor{DarkerGrey}{RGB}{215,217,219}
\definecolor{FontColor}{RGB}{131,136,138}
\definecolor{Red}{RGB}{204,0,0}
\definecolor{L-lig}{RGB}{25,124,192}
\definecolor{point-lig}{RGB}{54,104,163}
\definecolor{G-lig}{RGB}{62,66,68}

\definecolor{Orange}{RGB}{240,163,10} 
\definecolor{SCgreen}{RGB}{100,141,47}
\definecolor{Gray}{RGB}{186,200,211}
\definecolor{LightRed}{RGB}{214,98,93}
\definecolor{LightBlue}{RGB}{160,200,217}
\definecolor{LightGreen}{RGB}{130,161,119}
\definecolor{Violet}{RGB}{190,144,252}

\newcommand{\aj}[1]{a^{(#1)}}
\let\Tr\relax\DeclareMathOperator\Tr{\mathrm{Tr}}%

\let\Re\relax \DeclareMathOperator\Re{\mathrm{Re}}%

\usepackage{tikz}
\usetikzlibrary{arrows,shapes}
\usetikzlibrary{external}
\begin{document}

\author{Simon Kothe}
\affiliation{Department of Physics and SUPA, University of Strathclyde, Glasgow G4 0NG, United Kingdom}
\author{Peter Kirton}
\affiliation{Department of Physics and SUPA, University of Strathclyde, Glasgow G4 0NG, United Kingdom}
\date{\today}
	
\title{Liouville Space Neural Network Representation of Density Matrices}

\begin{abstract}

Neural network quantum states as ansatz wavefunctions have shown a lot of promise for finding the ground state of spin models. Recently, work has been focused on extending this idea to mixed states for simulating the dynamics of open systems.  Most approaches so far have used a purification ansatz where a copy of the system Hilbert space is added which when traced out gives the correct density matrix. Here, we instead present an extension of the Restricted Boltzmann Machine which directly represents the density matrix in Liouville space. This allows the compact representation of  states which appear in mean-field theory. We benchmark our approach on two different version of the dissipative transverse field Ising model which show our ansatz is able to compete with other state-of-the-art approaches.
	
\end{abstract}
	
\maketitle

\section{Introduction}

Developing techniques to study the many-body physics of open quantum systems opens up the avenue to access behavior beyond that possible in equilibrium.  This is relevant for understanding the dynamics of a variety of experimental platforms, but it also allows us to address fundamental questions about what kinds of physics is realizable in situations where incoherent driving and dissipation compete with coherent Hamiltonian dynamics. For example, in arrays of superconducting qubits coherent hopping of excitations can compete with onsite losses~\cite{Fitzpatrick2017}, or in semiconductor microcavities non-linearities in the Hamiltonian can compete with photon losses~\cite{Rodriguez2017}. 

Performing accurate simulations of the models describing this kind of physics can be very challenging. Even in the simplest case, where the dynamics of the system is well captured by a Markovian master equation~\cite{Breuer2002}, the effective size of the space required grows exponentially with the number of degrees of freedom included. This makes it difficult to make concrete statements about what phases are stable at large systems sizes, in the thermodynamic limit. Thus, developing new techniques, able to capture all of the required physical processes is a key challenge to be overcome. 

For closed systems a variety of techniques are available, allowing calculation of both ground states and dynamics in a variety of situations. These range from density functional theory, which relies on approximations of the electronic density~\cite{Jones2015}, to Monte Carlo methods~\cite{Metropolis1953,Hastings1970,Swindsen1986,Katzgraber2006,Katzgraber2011} which draw a small number of samples from the Hilbert space approximating the correct distribution, and tensor network methods~\cite{Vidal2003,Schollwoeck2011,Orus2014}, which rely on compression of the full many-body wavefunction. 

Recently, there has been much progress in using neural networks to accurately represent wavefunctions of many-body systems~\cite{Carleo2017, Carleo2019, Melko2019}. These neural network quantum states (NNQS) rely on the ability of neural networks to learn and represent any sufficiently smooth function~\cite{Kolmogorov1956,Hornik1991,Roux2008}, allowing an efficient description without requiring an exponential number of parameters. The simplest architecture used for NNQS is the Restricted Boltzmann Machine (RBM). They consist of two fully connected layers, one visible and one hidden. This simple structure of the architecture translates to a function which can be easily implemented numerically, allowing for an efficient Markov sampling of the relevant probability distribution ~\cite{Pei2021,Becca2017} and efficient optimization. By increasing the hidden unit density the expressibility of the ansatz increases, allowing them  to represent any state in the limit of infinite parameters. This has already shown to be competitive with other state-of-the-art approaches for finding ground~\cite{Carleo2017} and excited~\cite{Choo2018} states of 1D and 2D spin systems. This ansatz has been applied to a wide range of problems in this area, from simulating topologically complex states~\cite{Glasser2018} to investigating how the eigenvalue spectrum of the quantum Fisher matrix gives information about how networks learn groundstates~\cite{Park2020}. RBMs  are able to represent some states with a volume-law entropy scaling~\cite{Deng2017a}. 


Architectures beyond the RBM have also been used to study related problems. These range from recurrent neural networks~\cite{HibatAllah2020, Khandoker2022} to convolutional neural networks~\cite{Choo2019,Efthymiou2019,Roth2021} and deep autoregressive models, which forgo the need for Monte Carlo sampling altogether~\cite{Sharir2020}. 
 
These approaches have also begun to be applied to the dynamics and steady-states of open quantum systems. Neural density machines (NDMs)~\cite{Torlai2018, Hartmann2019, Nagy2019, Vicentini2019} use a pair of copies of a pure state NNQS coupled together by an extra layer of hidden neurons. The density matrix of interest is then  obtained by tracing out the degrees of freedom associated with one of the copies. Such purification schemes have been also used  in tensor network approaches~\cite{Nielsen2010,Bartel2009,Schollwoeck2011}. Other developments use a POVM representation of the density matrix, achieving accurate results~\cite{Carrasquilla2019, Reh2021, Luo2022} as well as autoregressive Gram-Hadamard density operators which extend the NDM by adding  additional layers which allows the capture of more complex correlations~\cite{Vicentini2022}.  Another ansatz was introduced by \citet{Yoshioka2019} which uses a binary encoding to vectorise their density matrix. So far, the literature has focused mostly on purifying the density matrix in Hilbert space, with the exception of  Ref.~\cite{Yoshioka2019}. Here, we propose a different approach, which writes the density matrix directly in Liouville space. The Liouville density machine (LDM) ansatz which we propose does not require additional visible units and maintains a correspondence of the visible layer and the concrete physical system. We show that this is able to efficiently represent a larger range of physically relevant states than the NDM improving learning and accuracy, while retaining the simplicity of RBMs.

This paper is organized as follows: In Sec.~\ref{sec_Markov} we give a brief description of the Lindblad master equation which describes Markovian open system dynamics. Sec.~\ref{sec_SS} details how finding the steady-state of this equation can be recast in terms minimization of a cost function and how it can be estimated using Monte Carlo sampling. Then, in section \ref{sec_NNQS}, we give a brief introduction to NNQS, using the original ansatz of Ref.~\cite{Carleo2017}, going on to show how this can be generalized to open systems. Sec.~\ref{sec_Res} gives a detailed comparison of the different approaches to solve this problem, using two versions of the dissipative transverse field Ising model as a benchmark. We gain insights into which kinds of states are easy and difficult to represent efficiently with the architectures described. Finally in Sec.~\ref{sec_Con} we give our conclusions and discuss possible ways to build upon these results. 
The appendices contain a detailed derivation of the stochastic reconfiguration algorithm and an explanation of how we produce the Markov chains for estimating the required expectation values from the neural network.

\section{Open System Dynamics} \label{sec_Markov}

The presence of an external environment fundamentally changes the nature of the dynamics of a quantum system. The state is no longer captured by a wavefunction and the non-unitary dynamics cannot be described by the Schr\"odinger equation.  Instead, the natural description is in terms of a reduced density operator for the system, $\rho$, and a master equation which defines its time evolution. This then allows for energy dissipation, decoherence and, most importantly for the present manuscript, the relaxation into a non-equilibrium steady-state in the long time limit.

If the coupling to the environment is weak and structureless, one can assume that system-induced correlations within the bath decay faster than the effects of the bath on the system. This leads to the particularly simple Lindblad master equation~\cite{Breuer2002} which takes the form

\begin{equation}\label{lindbladian}
\frac{d\rho}{dt}=\mathcal{L} \rho = -i[H,\rho] + \sum_i \gamma_i \mathcal{D}[A_i]\rho. 
\end{equation}
Here, the sum runs over the dissipation channels $i$ which are defined by the jump operators $A_i$ and rates $\gamma_i$. The dissipation superoperators are given by the Lindblad form 

\begin{equation}
	\mathcal{D}[A]\rho =  A \rho A^{\dagger} - \frac{1}{2} \left(A^{\dagger}A\rho+\rho A^{\dagger}A \right).
\end{equation}
We only consider time-independent master equations. 

The time evolution of the density operator is then governed by the formal expression 
\begin{equation}\label{time_evolution}
	\rho(t) = \text{e}^{\mathcal{L}t}\rho(0),
\end{equation}
such that in the long time limit the stationary state satisfies
\begin{equation}
	\lim_{t\to\infty}\mathcal{L} \rho(t) = 0.
\end{equation} 

To simplify what follows both mathematically and numerically we recast Eq.~\eqref{lindbladian} in Liouville space (sometimes also referred to as Choi's space). In this space the density matrix is reshaped into a vector,

\begin{equation} \label{dense_ket}
\rho= \sum_m \sum_n \rho_{m,n} \ket{m}\bra{n} \to \ket{\rho}\rangle = \sum_{m,n}\rho_{m,n}\ket{m,n}\rangle.
\end{equation}
Here the density matrix is expanded in the basis $\{\ket{m}\}$ with expansion coefficients $\rho_{m,n}$. The double-ket notation, $|\cdot\rangle\rangle$, used above denotes a vectorized operator which lives in a Hilbert space consisting of two copies of the original, $|\rho\rangle\rangle \in \mathcal{H}\otimes\mathcal{H}$. Then superoperators which act on these operator-kets are elements of Liouville space $L \in (\mathcal{H}\otimes\mathcal{H})^*\otimes (\mathcal{H}\otimes\mathcal{H})$~\cite{Minganti2018,Yoshioka2019}. Here we often abbreviate the double index $(m,n)$ with a single index $s$ which runs over the ket- and bra-indices of the density matrix $\sum_{m,n}\rho_{m,n}\ket{m,n}\rangle \equiv \sum_{s}\rho(s)\ket{s}\rangle$. With this Eq.~\eqref{lindbladian} now reads:

\begin{equation}
\frac{d}{d t}\ket{\rho}\rangle = L\ket{\rho}\rangle.
\end{equation}
where 
\begin{align*}
L = &-i\left(H\otimes\mathbb{1} - \mathbb{1}\otimes H^T  \right)\\ &+ \sum_k \left[A_k\otimes A_k^* - \frac{1}{2}\left(\mathbb{1}\otimes(A_k^{\dagger}A_k)^T + (A_k^{\dagger}A_k)\otimes\mathbb{1}\right)\right]
\end{align*}
is the matrix  form of the superoperator $\mathcal{L}$ that acts from the left on a density-ket, $\ket{\rho}\rangle$.

We can then use standard linear algebra results to find the eigenvalues and eigenkets of the matrix $L$, such that

\begin{equation}
	L\ket{\rho_i}\rangle = \lambda_i\ket{\rho_i}\rangle.
\end{equation}
A stationary state has $\lambda_i=0$ while all other states have $\Re \lambda_i<0$.

As we have seen above, the size of the required Liouville space scales much more quickly than the already exponentially growing Hilbert space. For example, for spin-1/2 lattice problems, the state vector of a closed system grows as $\mathcal{O}(2^N)$, with the system size $N$. The density matrix, on the other hand, grows as $\mathcal{O}(4^N$) and hence the Liouvillian has up to $\mathcal{O}(16^N)$ elements which limits the size of accessible systems even further. In the next sections we will detail our approach of tackling this problem.

\section{Finding the Steady-State} \label{sec_SS}

Many routes  to accessing larger system sizes with numerical simulations are variational methods~\cite{Weimer2015, Weimar2021}. For example, tensor network methods~\cite{Zwolak2004} can be seen as variationally finding the steady-state by optimizing over the set of states captured by a given tensor network architecture, while corner space RG~\cite{Finazzi2015} optimizes a particular (small) set of basis states which can be used to build up to larger system sizes. As we will see later, neural network techniques make use of the same underlying mathematical structures. Crucial to these methods is the use an ansatz function which specifies the full state with a finite number of parameters, which we denote by $\alpha$. We may then write $\rho(s;t) \to \rho_\alpha(s;t)$ and optimize over $\alpha$, to find the closest representation of the steady-state within this class. For this approach to be an efficient solution, the number of  parameters must grow at most polynomially with the system size.

To find the steady-state, $\lim_{t\to\infty}\rho(t) \equiv \rho_{ss}$, we require a cost function which measures how  close the ansatz is to the true stationary density operator. To do this we make use of the fact that $L\ket{\rho_{ss}}\rangle=0$ and that for all other states this quantity is finite. This gives us a cost function to optimize as well as a metric for how well the variational state parametrizes the true steady-state. The cost function we wish to optimize is given by the quantity

\begin{equation}\label{Cost}
|\text{C}(\alpha)|^2 = \left|\frac{\langle\langle \rho_{\alpha}|L|\rho_{\alpha}\rangle\rangle}{\langle\langle \rho_{\alpha}|\rho_{\alpha}\rangle\rangle}\right|^2,
\end{equation}
this uniquely vanishes in the steady-state. Note that the vectors appearing above  are in Liouville space, hence the denominator involves the sum over all elements of the density matrix ${\langle\langle \rho_{\alpha}|\rho_{\alpha}\rangle\rangle}=\sum_s |\rho_\alpha(s)|^2$. Expanding this cost function yields a form that can easily evaluated via Markov chain Monte Carlo

\begin{align*}
C_{\alpha} &= \frac{\langle\langle \rho_{\alpha}|L|\rho_{\alpha}\rangle\rangle}{\langle\langle \rho_{\alpha}|\rho_{\alpha}\rangle\rangle}\\
&=\frac{1}{\sum_{s'} |\rho_\alpha(s')|^2}\sum_{s,s'} \rho_{\alpha}^*(s)\rho_{\alpha}(s')\langle\langle s|L|s'\rangle\rangle\\
&=\frac{1}{\sum_{s'} |\rho_\alpha(s')|^2} \sum_{s,s'} \frac{|\rho_{\alpha}(s)|^2 \rho_{\alpha}(s')}{\rho_{\alpha}(s)}\langle\langle s|L|s'\rangle\rangle\\
&= \sum_s p(s;\alpha)  C_{\text{loc}}(s;\alpha),
\end{align*}
where 
\begin{equation}
	p(s;\alpha) = \frac{|\rho_\alpha(s)|^2}{\sum_{s'} |\rho_\alpha(s')|^2},
	\end{equation}
defines a probability distribution over the entire density matrix which we can draw samples from. The local cost associated to each element of this distribution is then 

\begin{equation}
C_{\text{loc}}(s;\alpha)= \sum_{s'}\langle\langle s|L|s'\rangle\rangle\frac{\rho_{\alpha}(s')}{\rho_{\alpha}(s)}.
\end{equation}
 All parts of this local cost can be efficiently calculated. The states $s'$ which connect, via the matrix element of the Liouvillian, to the original state $s$  are generated during the Metropolis-Hastings step and can be accessed at any later stage of the process, see App.~\ref{MCMC} for details on the algorithm used. Since, for a local Liouvillian, the number of non-zero elements grows only linearly in system size, the evaluation of the matrix elements of $L$ can also be done efficiently. The samples we draw follow the distribution $p(s;\alpha)$, and so the cost can be calculated as a simple mean over the local costs~\cite{Becca2017}

\begin{equation}
C_{\alpha} \approx \frac{1}{N_s}\sum_s C_{\text{loc}}(s;\alpha),
\end{equation}
where $N_s$ is the number of samples and  the sum runs over the Monte Carlo samples. 

Along with an estimate for the cost function we also need a way of updating the parameters, $\alpha$ such that the state we find is optimized.  The simplest way to do this is via stochastic gradient ascent (SGA), where the gradients of $C_{\alpha}$ are also estimated with the Monte Carlo samples and at each step in the simulation the parameters are updated as

\begin{equation}
\alpha \rightarrow \alpha' = \alpha + \eta \nabla_{\alpha} C_{\alpha},
\end{equation}
with the learning rate $\eta$. There are several problems with this approach, e.g.~it has been shown~\cite{Park2020} that SGA has severe problems with steep energy surfaces.  The main problem for the present case is that $L$  also has a left eigenstate with eigenvalue 0,

\begin{equation}
\langle\bra{\mathbb{T}}L = 0.
\end{equation}
This is the trace-state which is defined as 

\begin{align}
\langle\langle \mathbb{T}|\rho\rangle\rangle &= \Tr\left[\rho\right],
\end{align}
and since the dynamics of any physical master equation is necessarily trace preserving we find the result above.
This means that there is another state which optimizes the cost function.
A solution to both of these problems is to instead use {Stochastic Reconfiguration} (SR)~\cite{Sorella1998, Sorella2001, Becca2017, Park2020} to update the parameters. SR can be derived by asking which parameter update $\gamma$,
\begin{equation}
 \alpha \rightarrow \alpha' = \alpha + \eta \gamma,
\end{equation} 
best approximates a step in real time (see App.~\ref{app:SR} for a derivation). Since the trace state cannot be found by a real-time evolution generated by $L$ this guarantees that we will find the correct steady-state when optimizing the cost, Eq.~\eqref{Cost}. Furthermore, SR takes into account the curvature of the energy landscape, speeding up the optimization on flat areas and slowing down in the presence of strong curvature. The result is that the updates are calculated as

\begin{equation}
\gamma = S^{-1} f,
\end{equation}
where $S_{i,j} = \langle O_i^*O_j \rangle - \langle O^*_i\rangle\langle O_j\rangle$ is the quantum Fisher matrix and $f_i = \langle O^*_i L\rangle - \langle O^*_i\rangle \langle  L\rangle$ is the the gradient of the cost function in Eq.~\eqref{Cost}. The angle brackets, $\langle \ \cdot\ \rangle$, denote the expectation value over the Monte Carlo samples and the operator $O_i(s)$ is the logarithmic derivative of $\rho_{\alpha}(s)$ with respect to the $i$th parameter

\begin{equation}
O_i(s) = \frac{1}{\rho_{\alpha}(s)}\frac{\partial}{\partial \alpha_i} \rho_{\alpha}(s).
\end{equation}
The matrix $S$ is defined to be positive, however when estimating it via Monte Carlo sampling it can happen that some eigenvalues vanish and $S$ becomes singular. One can work around this problem by either calculating the pseudo-inverse or adding a small regularization, $\lambda \approx 10^{-3}$, to the diagonal. We employ the latter method here. 

We next need to specify the form of our ansatz function for the density operator, $\rho_\alpha$. To do this we first give a brief overview of how neural networks can be used to represent many-body wavefunctions.

\section{Neural Network Quantum States} \label{sec_NNQS}

Neural network quantum states are a variational ansatz whose parameters and functional form are defined by an underlying neural network architecture. This approach has been recently developed to describe the many-body wavefunction of interacting spin systems and was introduced in Ref.~\cite{Carleo2017}. The key insight here is that the heart of the many-body problem is to find the relevant parts of Hilbert space in which e.g.~the ground state of a system lives. This is essentially a problem of dimensionality reduction and feature extraction, which are the two strongest points of neural networks. 

\begin{figure}
       \includegraphics{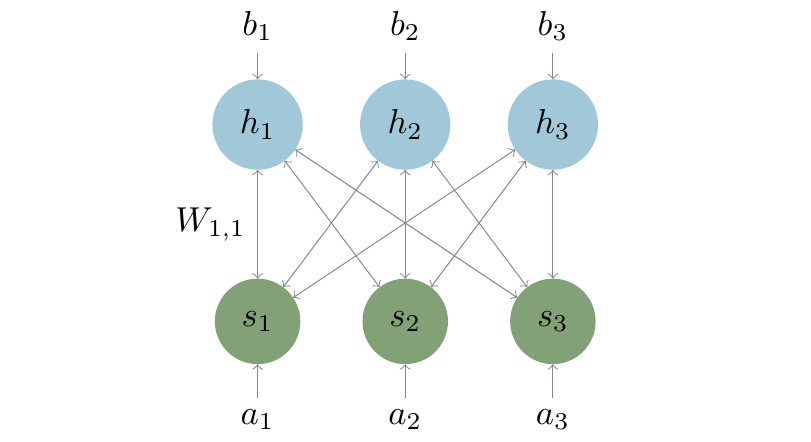}
       \caption{\label{fig:RBM} Schematic representation of a RBM. It shows the visible units (lower row, green) connected to the hidden units (upper row, blue) via the weight matrix (double-headed arrows). The single-headed arrows labeled by $a_i$ and $b_i$ show the biases applied to the neurons.}
\end{figure}

NNQS use this ability of neural networks to efficiently represent any continuous function~\cite{Kolmogorov1956,Hornik1991,Roux2008}, e.g.~a wavefunction~\cite{Carleo2017, Choo2018, Glasser2018} or density matrix~\cite{Hartmann2019, Nagy2019, Vicentini2019} by learning the most important features which define that function. 

The simplest NNQS architecture is the RBM. These are bilayer neural networks with one visible and one hidden layer of neurons. The network is parametrized by a bias attached to each neuron and a set of weights which connects the two layers. This setup is shown schematically in Fig.~\ref{fig:RBM}. In the same spirit as the discussion in the previous section, it is possible to then  write a variational ansatz for a wavefunction for a spin-1/2 lattice model. This takes the form

\begin{align}\label{pure}
\Psi_{\alpha}(s) &= \sum_{\{h_i\}} \text{e}^{\sum_j a_j s_j + \sum_i b_i h_i + \sum_{ij} W_{ij} h_i s_j}\\
&= \text{e}^{\sum_j a_j s_j}\prod^M_{i=1} 2 \cosh\left[b_i + \sum_j^N W_{ij} s_j\right].
\end{align}
Where in the second line we use the fact that the sum over configurations can be analytically calculated. Here $a_i(b_j)$ denotes the visible (hidden) bias of the $i$th  site($j$th hidden unit), and $W_{ij}$ the weights connecting the $j$th visible unit to the $i$th hidden unit. $M(N)$ denotes the number of hidden(visible) units. The ratio between hidden and visible units is $\beta = M/N$. For $\beta=0$ the only states which can be described are those that would arise in a mean-field description, i.e.~those which are a product of single site wavefunctions. Increasing the value of $\beta$ allows a systematic increase in the amount of entanglement which can be described~\cite{Carleo2017}. A further strength of the RBM wavefunction is the simplicity of its derivatives with respect to the parameters. This allows for efficient evaluation of all of the derivatives required to update the state. 


In its simplest form the RBM ansatz described above  is only able to discriminate betwen two different visible states, $s = [-1,1]$, which allows at most for the representation of pure states of  spin-$1/2$ systems.  However, this is not sufficient to describe a vectorised density operator in Liouville space which, even for a spin-$1/2$ system has 4 elements. To overcome this issue we use an approach based on that introduced in Ref.~\cite{Pei2021} for the study of pure states of spin-1 lattices. This was achieved by adding an additional set of biases and weights which allows discrimination between the $s=[-1,0,1]$ states. 

\begin{figure}[h] 
\includegraphics{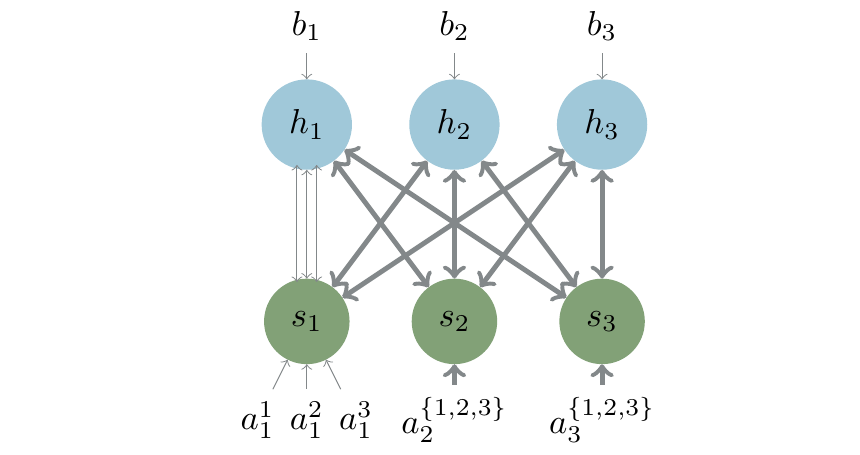}
\caption{\label{fig:LDM} Schematic representation of an LDM. Similar to the RBM, it has visible (lower row, green) and hidden (upper row, blue) units. LDMs, however have three visible biases $a^{\{1,2,3\}}_i$ as well as three weight matrices $U$, $V$, and $W$ (double arrows). The weight matrices couple to different powers of the configuration vector $s$ and encode correlations between different local states.}
\end{figure}

We can adapt this idea to provide a basis for a NNQS ansatz for density-kets, $|\rho\rangle\rangle$. Such an ansatz needs to be able to discriminate between the four states of the local density operator, while maintaining a one-to-one correspondence between the visible layer and the physical system. To label the density matrix elements on each site we choose to use the mapping
 \begin{equation}
	s = \begin{cases}
		\hphantom{-}2 \to \ket{\uparrow\uparrow}\rangle\\
		\hphantom{-} 1  \to \ket{\uparrow\downarrow}\rangle\\
		-1 \to \ket{\downarrow\uparrow}\rangle\\
		-2 \to \ket{\downarrow\downarrow}\rangle
	\end{cases}
\end{equation}
where $s=\pm 2$ denote the diagonal density matrix elements, while $s=\pm 1$ are the off-diagonal elements. These values are chosen as they give a simple mapping from integers to density matrix elements and allow our neural network ansatz to easily distinguish between all of the states. To discriminate these four different local states  we add one more set of visible biases and weights, allowing us to represent a density-ket. Furthermore, we do not require any additional visible nodes to achieve this, as in Ref.~\cite{Yoshioka2019}. The total ansatz function now reads:

	\begin{multline}
		\rho_{\alpha}(s) = \text{e}^{\sum_j^N \aj{1}_j s_j + \aj{2}_j s_j^2 + \aj{3}_j s_j^3}\\
		\times\prod^M_{i=1} 2 \cosh\left[b_i + \sum^N_j U_{i,j} s_j + V_{i,j} s_j^2 + W_{i,j} s_j^3\right],
	\end{multline}
where $N$ and $M$ are the number of visible and hidden nodes respectively.   We have introduced a set of complex biases, $\aj{n}$ to the visible units which allow us to distinguish different visible biases, $b$ again gives the bias on each hidden neuron and $U$, $V$, $W$ are the complex weight matrices that connect the two layers. Theses biases and connections are shown schematically in Fig.~\ref{fig:LDM}. Similar to the pure-state case of Eq.~\eqref{pure}, this ansatz has labeling freedom and very simple derivatives which can be found analytically and efficiently implemented. Calculating the derivatives of a complex valued and complex parametrized function is done using Wirtinger calculus~\cite{Amin2012}. 

{ If the hidden unit density, $M = 0$, this ansatz is able to exactly capture those states described by  mean-field theory, i.e.~where the full density matrix can be factorized onto individual sites. This then allows the parameter $M$ to systematically increase the amount the correlations which can be represented in a  similar way to the bond dimension of an MPS. }

{This ansatz, which we refer to as a \textit{Liouville Density Machine} (LDM), bears some similarities to that proposed in Ref.~\cite{Yoshioka2019}. In both cases the problem is transformed into Liouville space but in Ref.~\cite{Yoshioka2019} this is achieved by adding an additional visible unit on each site,  allowing them to represent the four required states. A detailed comparison between the expressibility of these approaches is left as an open question. The LDM ansatz should be seen in contrast to those in Refs.~\cite{Torlai2018, Vicentini2019, Hartmann2019, Nagy2019}.} In these papers a purification ansatz is used, an extra layer of hidden units is added to represent the state of an auxiliary system which, when traced over, leaves the appropriate density matrix for the original system.  Similar techniques have been applied to MPS simulations~\cite{Nielsen2010,Bartel2009}. In what follows we will refer to this purification ansatz as a \textit{Neural Density Machine} (NDM).

Compared to the purification approaches of Refs.~\cite{Torlai2018, Vicentini2019, Hartmann2019, Nagy2019}, the LDM is not guaranteed by construction to be either Hermitian or positive-definite. However, we find that this is not a significant issue in our simulations because of our choice of SR as the descent algorithm. Note that any matrix can be represented in the eigenbasis of the Liouvillian~\citep{Minganti2018} as
\begin{equation}
\ket{A}\rangle =c_0 \ket{\rho_0}\rangle + \sum_{i\neq0} c_i \ket{\rho_i}\rangle,
\end{equation}
where $c_i$ are expansion coefficients and $\ket{\rho_i}\rangle$ are the vectorized eigenmatrices of $L$. Under real-time evolution, all eigenvalues of eigenstates except for the steady state, $\ket{\rho_0}\rangle$, have a real-part $\text{Re}(\lambda_{i\neq 0})< 0$ and hence exponentially decay over time. Hence, \textit{any} initial matrix, independent of its physicality, must necessarily converge to the steady-state of the Liouvillian. We stress now, that up to first order, the optimization algorithm we employ was derived as an approximation to real-time evolution under a Liouvillian and hence gives a good approximation to this kind of behavior.

\section{Results} \label{sec_Res}

To benchmark and understand the strengths and limitations of  the LDM ansatz we study the stationary state of the 1D dissipative transverse-field Ising (TFI) model~\cite{Bardyn2012, Lee2013, Joshi2013,Yoshioka2019, Luo2022, Vicentini2019}.  The system consists of a chain of spin-1/2 particles. The Hamiltonian part of the evolution is governed by
\begin{equation}
\label{TFI_z}
H = \frac{J}{4}\sum_{i}^{N-1} \sigma^z_i\sigma^z_{i+1} + \frac{h}{2}\sum_{i}^N \sigma^x_i.
\end{equation}
Here, $N$ is the number of sites, $\sigma$ are the usual Pauli matrices, $J$ is the interaction strength and $h$ is the strength of the transverse field. The dissipation is governed by excitation loss on each site, so the jump operators which appear in the Liouvillian are $A_i = \sqrt{\gamma}\sigma^-_i$. We choose the units of the interaction and field such that the dissipation strength is $\gamma = 1$. In all calculations that follow we set the interaction strength to $J/\gamma=2$.

The steady-state of this model then has simple solutions in two limiting cases. When $h\to0$ the dissipation dominates the dynamics and the stationary state is a pure product-state with all the spins pointing down 
\begin{equation} \label{eqn:pureh0}
	\lim_{h\to 0} \rho_{ss} = \bigotimes^{N} \ket{\downarrow}\bra{\downarrow}.
\end{equation}
In the opposite limit where $h\to\infty$ there is only competition between the local onsite field and the dissipation and so the steady-state ends up again as a product-state, but this time the state on each site is  mixed
\begin{equation} \label{eqn:mixedhinf}
	\lim_{h\to \infty} \rho_{ss} = \bigotimes^{N} \frac{1}{2}\left(\ket{\uparrow}\bra{\uparrow}+\ket{\downarrow}\bra{\downarrow}\right).
\end{equation}
At intermediate values of $h$ the steady-state interpolates between these two, building up complex long range classical and quantum correlations.

\begin{figure}
	
	\centering
	\includegraphics[width=0.5\textwidth,scale=0.1]{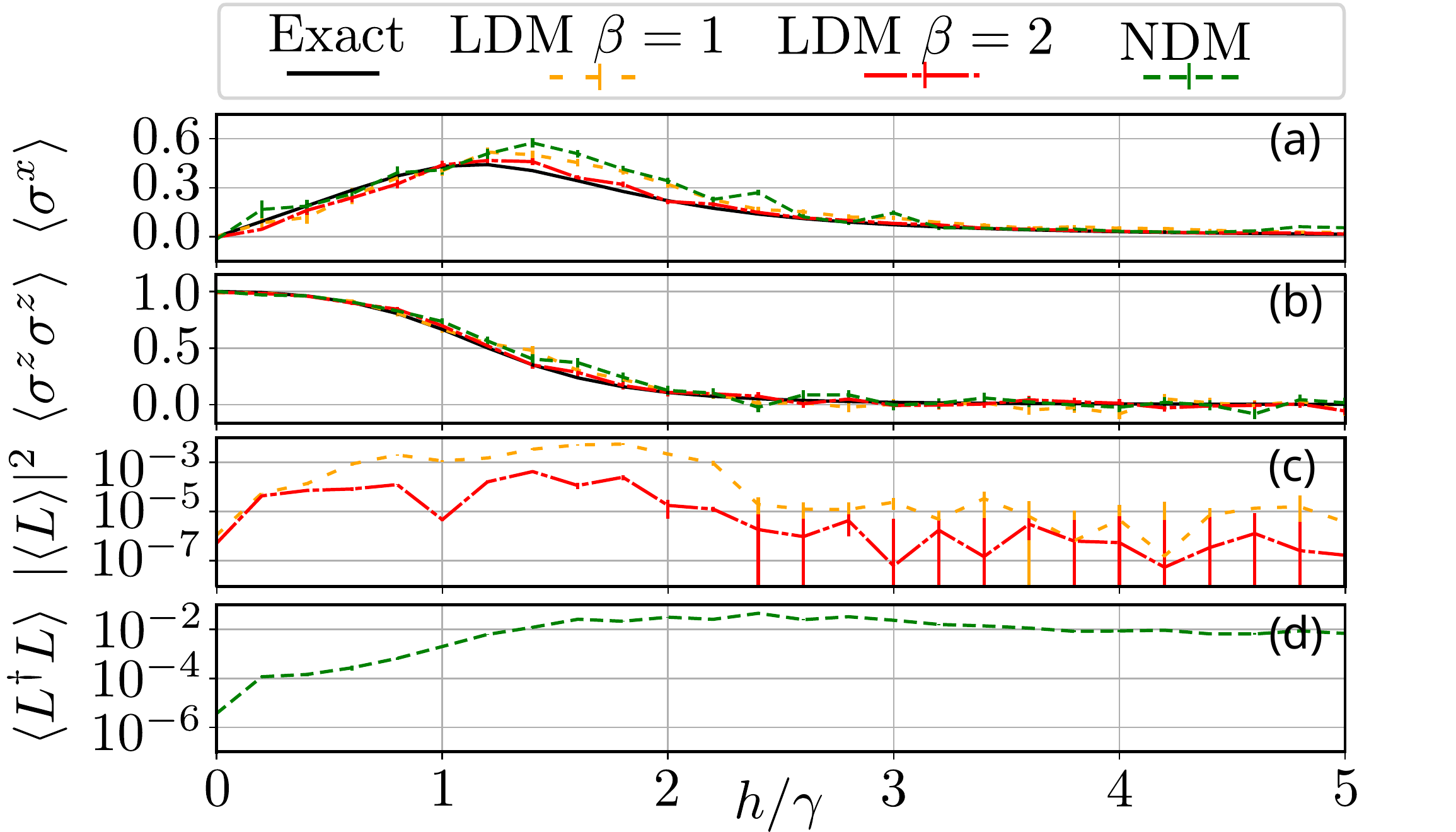}
	\caption{Comparison of the steady-state of Eq.~\eqref{TFI_z} using both the NDM and LDM approaches. The system size, $N=6$, is small enough that exact diagonalization is possible.  (a)~The expectation value of $\sigma^x_i$ as a function of $h$. (b)~The expectation value of the $\sigma^z_i\sigma^z_{i+1}$ correlation function. In both cases the expectation value is taken on the central site(s). The black (solid) lines show the exact result, the NDM with $\beta=1$ is the green (dashed) lines, while the LDM with $\beta =1$ is orange (doted) and with $\beta=2$ is red (dot-dashed). (c)~The absolute value squared of the cost function in Eq.~\ref{Cost} for the two LDM results. (d)~The expectation value $\langle L^\dagger L\rangle$ used as a cost function for the NDM ansatz employed by NetKet. For both cases with $\beta=1$ we used 4500 samples and optimized for 1000 steps with a learning rate of $\eta = 10^{-2}$ and a regularization of $\lambda = 10^{-2}$, for the expectation values we used 500 diagonal samples. In the $\beta=2$ case we used 6000 samples and 2000 steps and 800 diagonal samples. The other meta parameters were the same in both cases.}
	\label{Cost_z_fig}
\end{figure}

We will compare the results of the LDM ansatz to those obtained using the NDM approach as implemented in the NetKet library~\cite{Netket}. For small system sizes we will also be able to compare to results obtained from exact diagonalization, these are calculated using QuTIP~\cite{qutip}.

As a first example we look at the case of a small system with $N=6$. Our findings are summarized in Fig.~\ref{Cost_z_fig}. In each case we randomly initialize the parameter values and at each step we take Monte Carlo samples to approximate the cost function and the best updates to the parameters. At the end of each run we produce a new Markov chain, but this time sample from a probability distribution which  follows the diagonal of the density matrix. This allows us to estimate the expectation value of observables of interest. Since there are fewer diagonal states than entries in the full density matrix we usually only need about 500-800 diagonal samples.  

In panels (a)--(b) of Fig.~\ref{Cost_z_fig} we show the expectation value of $\sigma^x_i$ on the central site and the $\sigma^z_i\sigma^z_{i+1}$ correlation function on the central pair of sites as a function of the field strength, $h$. This gives a good indication of how well the various approaches are able to produce single site observables. We see that, in general, there is good agreement between the exact results and those obtained using both the LDM and NDM approaches. In all cases the agreement is worst in the central region where $1 \leq h/\gamma \leq 2.5 $, which is in agreement with the results of Ref.~\cite{Vicentini2019}. For the LDM a hidden unit density of $\beta=1$ corresponds to 132 parameters for the NDM this is 174 parameters. We see that even with only $3/4$ of the parameters the LDM generally gives as good or better results than the NDM.  By increasing the number of hidden units in the LDM we also increase the number of parameters so that, at $\beta=2$, there are 246 parameters. We see that for the LDM ansatz increasing the value of $\beta$ and hence the number of parameters used is able to significantly decrease the deviation from the exact result, thus we may use the hidden unit density as a way of checking for convergence when exact results are no longer possible. We can also see this effect more clearly in Fig.~\ref{Cost_z_fig}(c) and (d) where we show the Monte Carlo estimated cost function for each ansatz. The value of the cost function is significantly decreased at all values of $h$ when $\beta$ is increased. We also see that for the LDM ansatz the cost function is at a maximum in the regions where the convergence to the steady-state is worst, this allows us to use this estimate to again judge the accuracy of our results for system sizes where exact methods are unavailable. This is not true of the NDM approach where the cost function reaches a maximum at intermediate values of $h$ and does not significantly decrease as $h$ increases further.  This is because the mixed product state, described in Eq.~\eqref{eqn:mixedhinf}, is not so easy to represent in a purification ansatz; this mixed state requires a large amount of entanglement between the real and auxiliary spins. This is not the case for the LDM approach which can represent this state exactly without using hidden units.

\begin{figure}
	\centering
	\includegraphics[width=0.5\textwidth,scale=0.1]{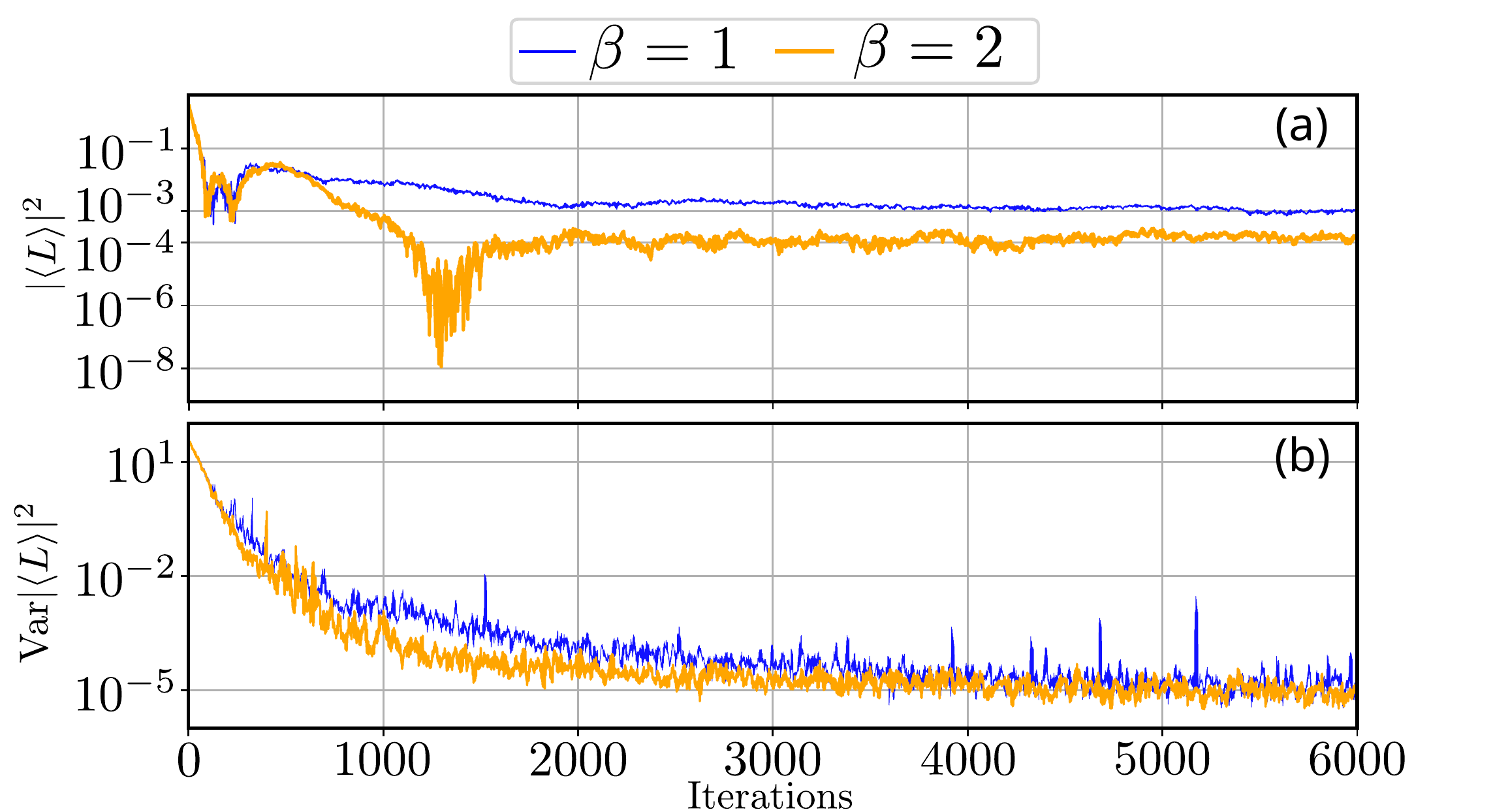}
	\caption{Convergence of the LDM ansatz for two different hidden unit densities, $\beta=1$ (thin, blue line) and $\beta=2$ (orange, thick line) . Panel (a) shows the running estimate for the cost function while panel (b) is the variance in the same quantity.  Increasing the hidden unit density improves the accuracy of the results. Both calculations were done at $h/\gamma=1$ using the same parameters as Fig.~\ref{Cost_z_fig}.
	}
	\label{convergence}
\end{figure}

A further comparison is shown in Fig.~\ref{convergence}. This figure shows how the cost function of the LDM evolves for two different values of $\beta$ over 6000 steps at one of the most difficult points, $h=\gamma$. By increasing the number of variational parameters from $132$ to $246$ we were able to reduce the cost function by an order of magnitude. We also see that simply checking the value of the cost function does not give an accurate stopping condition for the algorithm. After around 1400 steps the cost function for $\beta=2$ is very small but the variance is quite large. This means that the LDM has not found an eigenstate of the Liouvillian but is still giving a small value for the  cost function. We propose that a condition based on a combination of both of these quantities can give a good way to automatically stop the learning process when a good approximation to the steady-state has been reached.

\begin{figure}
	
	\centering
	\includegraphics[width=0.5\textwidth,scale=0.1]{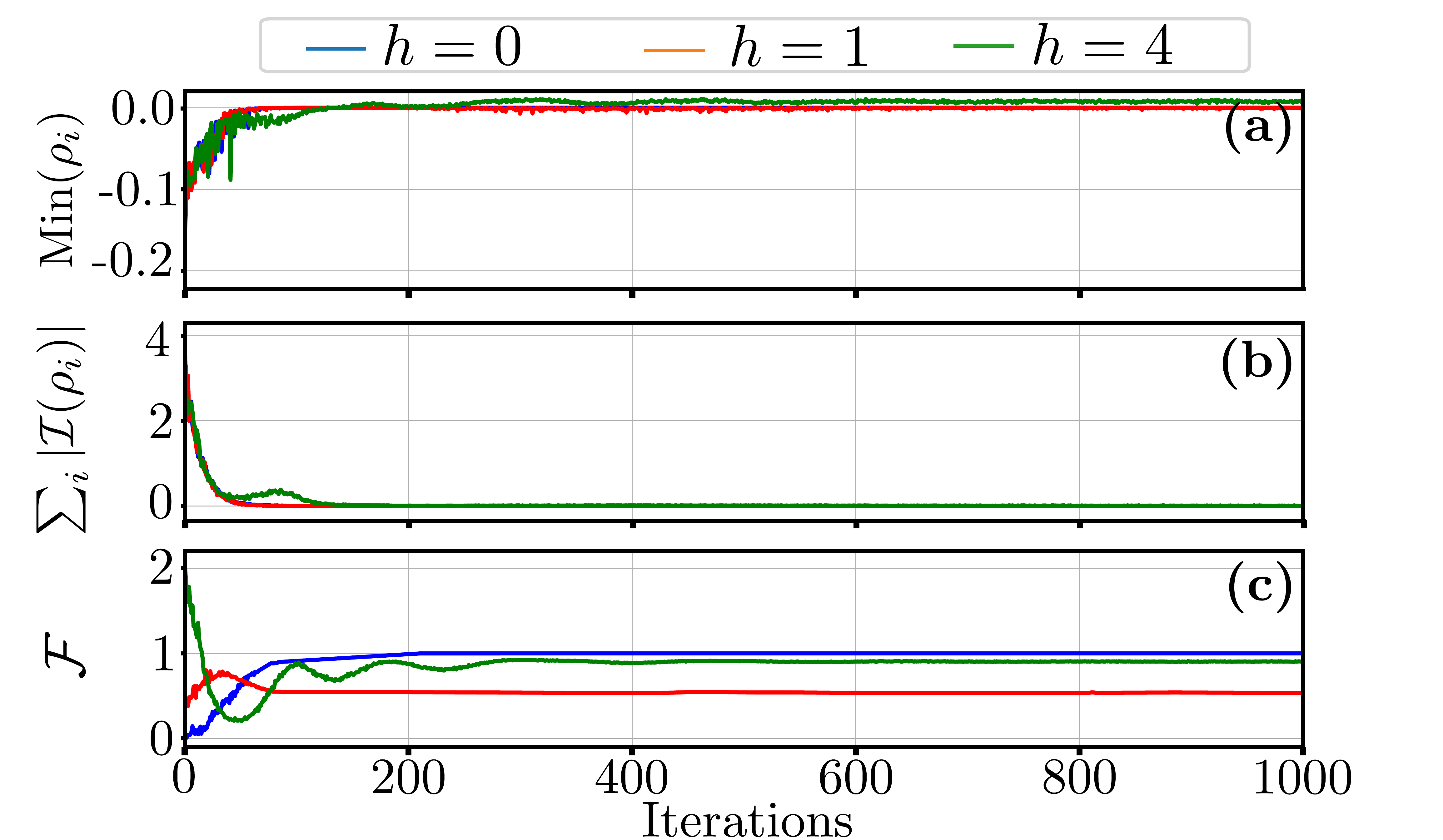}
	\caption{Properties of the state obtained by using SR at three different values of $h$ and all other parameters the same as the $\beta=1$ results in Fig.~\ref{Cost_z_fig}. Panel (a) shows the real part of the minimum eigenvalue of the full density matrix obtained from the LDM.  In (b) we show the sum of the absolute value of the imaginary parts of the eigenvalues of the density matrix. Panel (c) gives the fidelity of the ansatz with the exact density matrix. }
	\label{Min_e}
\end{figure}

To test that the state we obtain is physical we construct the full density matrix from the LDM. In the top panel of Fig.~\ref{Min_e} we show how the real part of the smallest eigenvalue of this constructed density matrix evolves. For a randomly initiated state the minimum eigenvalue is negative, which indicates a non-physical density matrix. However, as the optimization goes on the minimum eigenvalues become closer to 0 or becomes positive, indicating that the final density matrix is positive semi-definite. The middle panel of this figure shows the imaginary parts of all of the eigenvalues are quickly suppressed which indicates a hermitian matrix and hence we can be we have produced a physical density matrix. Inf Fig.~\ref{Min_e}(c) we show the fidelity of the state with one obtained from the exact solution of the master equation. We see that, as expected from the results of Fig.~\ref{Cost_z_fig}, the fidelity is highest in the cases where the transverse field strength is either very large or very small and the fidelity is lowest where the steady state has large entanglement around $h=1\gamma$. 

\begin{figure}	
	\centering
	\includegraphics[width=0.5\textwidth,scale=0.1]{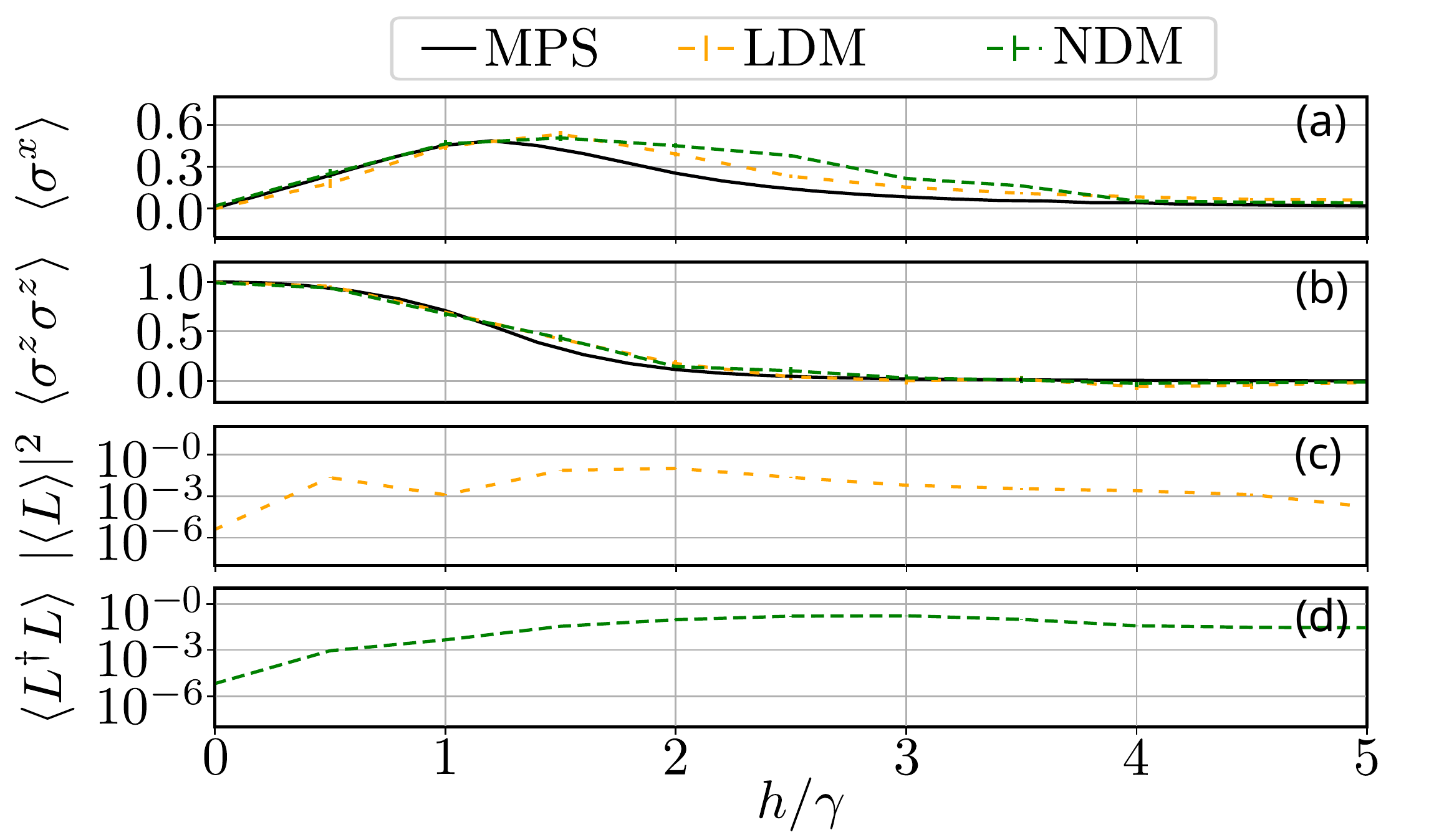}
	\caption{Steady-state of Eq.~\eqref{TFI_z}  as a function of  field strength, similar to Fig.~\ref{Cost_z_fig}, for a larger system size $N=16$. The expectation value of $\sigma^x_i$ and $\sigma^z_i\sigma^z_{i+1}$ on  the central site(s) are shown in panels (a)--(b) and the relevant cost functions in panels (c)--(d). The LDM (orange, dotted lines) and NDM (green, dashed lines) results are compared to those obtained from MPS simulations (black, solid lines).  We used $\beta=1.4$ in the case of LDM and $\beta=1$ for the NDM to ensure that both approaches use a similar number of parameters. The NDM has $1104$ parameters, while the LDM has $1126$. In both cases we evolved for 7000 steps with a learning rate of $\eta=10^{-3}$ we took $9000$ Monte Carlo samples at each step and a regularization of $\lambda = 3\times 10^{-3}$. We used 800 diagonal samples to estimate the expectation values.}
	\label{Cost_16}
\end{figure}

\begin{figure}
	\centering
	\includegraphics[width=0.5\textwidth,scale=0.1]{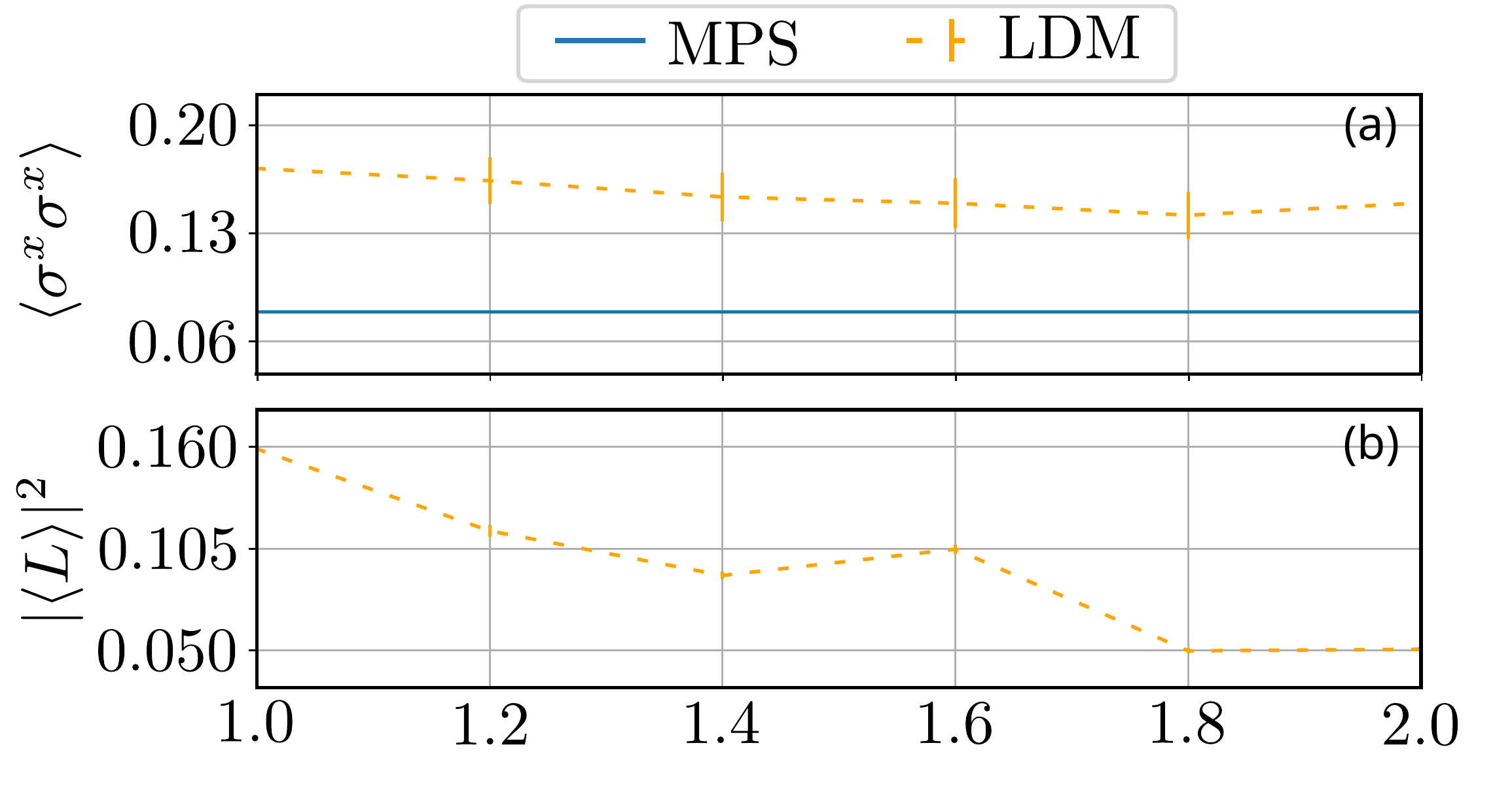}
	\caption{Improvement of convergence as a function of $\beta$ for the $N=16$ TFI model at $h=2\gamma$ (orange, dotted line) compared to the results of an MPS calculation (blue, solid line). In (a) we show the $\sigma_i^x\sigma_{i+1}^x$ correlation function and in panel (b) the estimate for the cost function. All calculations ran for 7000 steps. To accommodate for higher parameter counts we increased the number of samples with $\beta$ from $9000$ at $\beta=1$ to $17000$ at $\beta=2$. Other parameters are the same as in Fig.~\ref{Cost_16}. }
	\label{beta_scan}
\end{figure}

We now go on to examine how the accuracy of these approaches scales to larger system sizes. At $N=16$ it becomes difficult to use exact methods to compare against, however this model is straightforward to solve with MPS simulations which we found to be fully converged for a bond-dimension of $\chi = 7$. Results of these calculations are shown in Fig.~\ref{Cost_16}. The number of parameters for the NDM is $1104$ while the LDM has $1126$.  For reference, a bond dimension of $\chi=7$ corresponds to 1952 matrix elements in the MPS. We see very similar behavior to the $N=6$ case, both approaches are more difficult to converge in the region of intermediate $h/\gamma$ and the cost function for the LDM has a peak in this region. In Fig.~\ref{beta_scan} we show how the convergence can again be improved by increasing the hidden unit density. Here we choose $h=2\gamma$ as this is the point where the convergence is worst. We see that as $\beta$ is increased the cost function decreases towards zero and the expectation value moves towards that found in the MPS simulation. The expectation value here is a two-point correlation function which, in general, are harder to converge than single-site operators.

\subsection{TFI Model with rotated Hamiltonian}

By making a simple change to the model discussed above it is possible to make the convergence of both neural network approaches considerably worse. To do this  we change the Hamiltonian to a rotated basis~\cite{Bardyn2012,Joshi2013,Jin2016}

\begin{equation}
\label{TFI_x}
H = \frac{J}{4}\sum_{i=0}^{4} \sigma^x_i\sigma^x_{i+1} + \frac{h}{2}\sum_{i=0}^5 \sigma^z_i,
\end{equation}
but keep the dissipation processes the same as for the previous model. In this case the dissipation does not explicitly break the $\mathbb{Z}_2$ symmetry of the model as the interaction term is perpendicular to the dissipation. Therefore the competition between the coherent and dissipative dynamics gives rise to complex correlations in the steady-state. This leads to a very  rich mean-field phase diagram in high dimensions with possibilities for both first and second order phase transitions between different magnetic orderings~\cite{Lee2013, Mascarenhas2015, Jin2016}. In 1D these phase transitions turn into continuous crossovers, but complex correlations still build up when $h\sim J\sim \gamma$.  For a detailed review of the behavior of this model in 1D see Ref.~\cite{Joshi2013}.

\begin{figure}
	
	\centering
	\includegraphics[width=0.5\textwidth,scale=0.1]{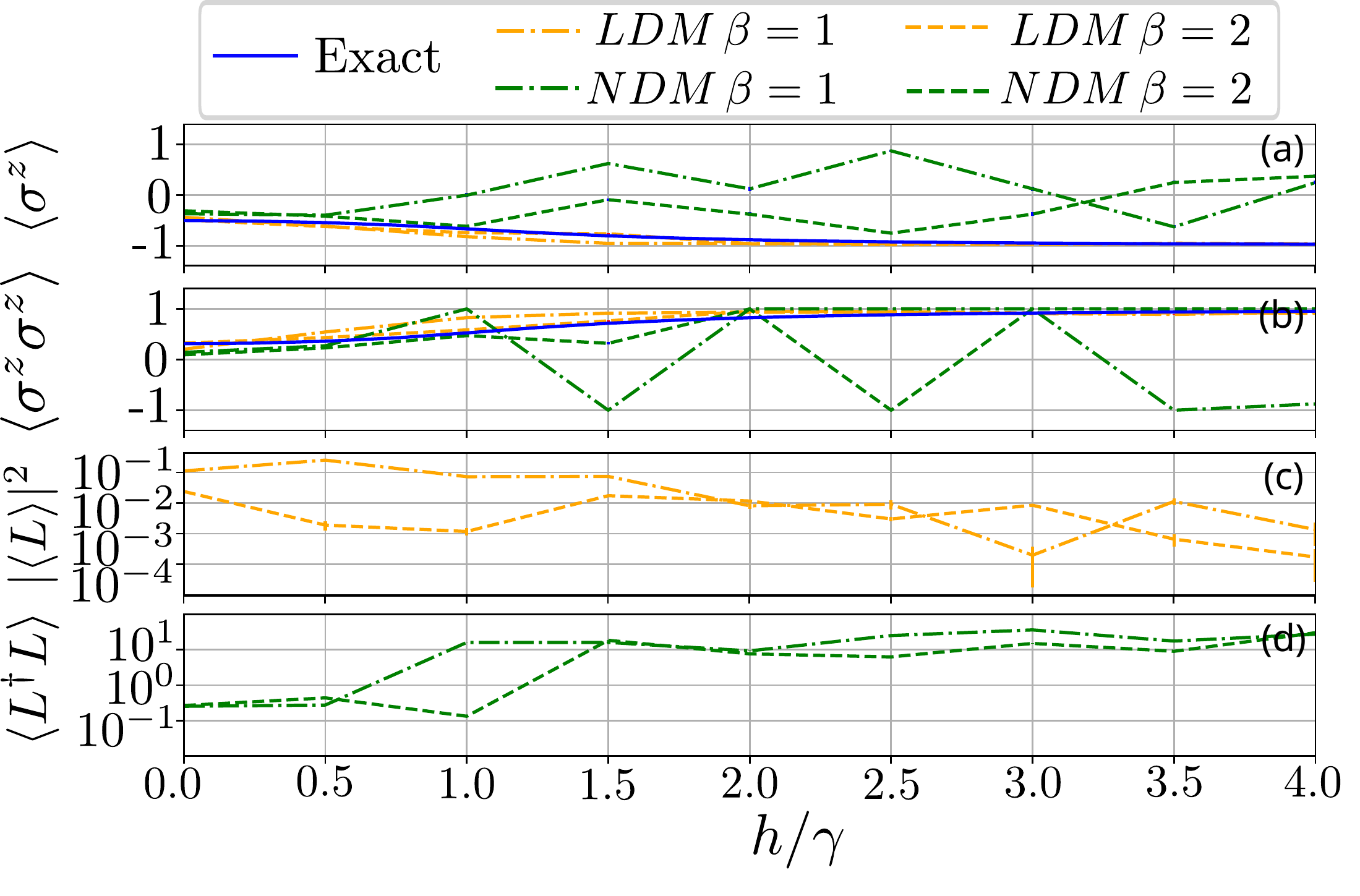}
	\caption{Optimizing the rotated TFI model as in Eq.~\ref{TFI_x} for $N=6$. The exact results are in blue, those obtained with the LDM are orange and the NDM  in green. Different hidden unit densities are shown by different line-styles. In all cases the optimization was run for 4000 steps, with a learning rate of $\eta=10^{-3}$ and $2000$ diagonal samples to estimate the expectation values. For $\beta=1$ we used $4500$ samples and a regularization of $\lambda= 10^{-3}$. For $\beta=2$ the number of samples was increased to $18000$ and the regularization was $10^{-2}$. In panel~(a) we give the steady-state expectation value of $\sigma^z_i$ and in (b) we show the two-point correlation function $\langle\sigma^z_i\sigma^z_{i+1}\rangle$. Panels (c) and (d) show the relevant cost functions for each ansatz. 
}
	\label{Cost_x_fig}
\end{figure}

We again study the convergence of both the LDM and NDM approaches for finding the steady-state of this model for a system size of $N=6$.  In panels (a) and (b) of Fig.~\ref{Cost_x_fig} we show how both a single site  and two-site observable varies with the applied field $h$. We see that, even when using a large amount of samples and parameters the NDM ansatz is not able to find a good approximation to the exact result, while the LDM is able to get much closer to the expected result, especially at small values of $h$. We see that for both approaches the cost function estimate is much larger than it was for the simpler model described by Eq.~\eqref{TFI_z}. This is because the steady-state in this case has much more complex correlations than in the previous model, simple expressions like those in Eqs.~\eqref{eqn:pureh0}--\eqref{eqn:mixedhinf} are not available, except at very large $h\to\infty$ where the steady-state is the same as given in Eq.~\eqref{eqn:pureh0}.  We next go on to show how using measures of the entanglement found in the steady-state can give good intuition for when these kinds of difficulties arise.

\subsection{Entanglement Properties}

\begin{figure}
	\centering
	\includegraphics[width=0.5\textwidth,scale=0.1]{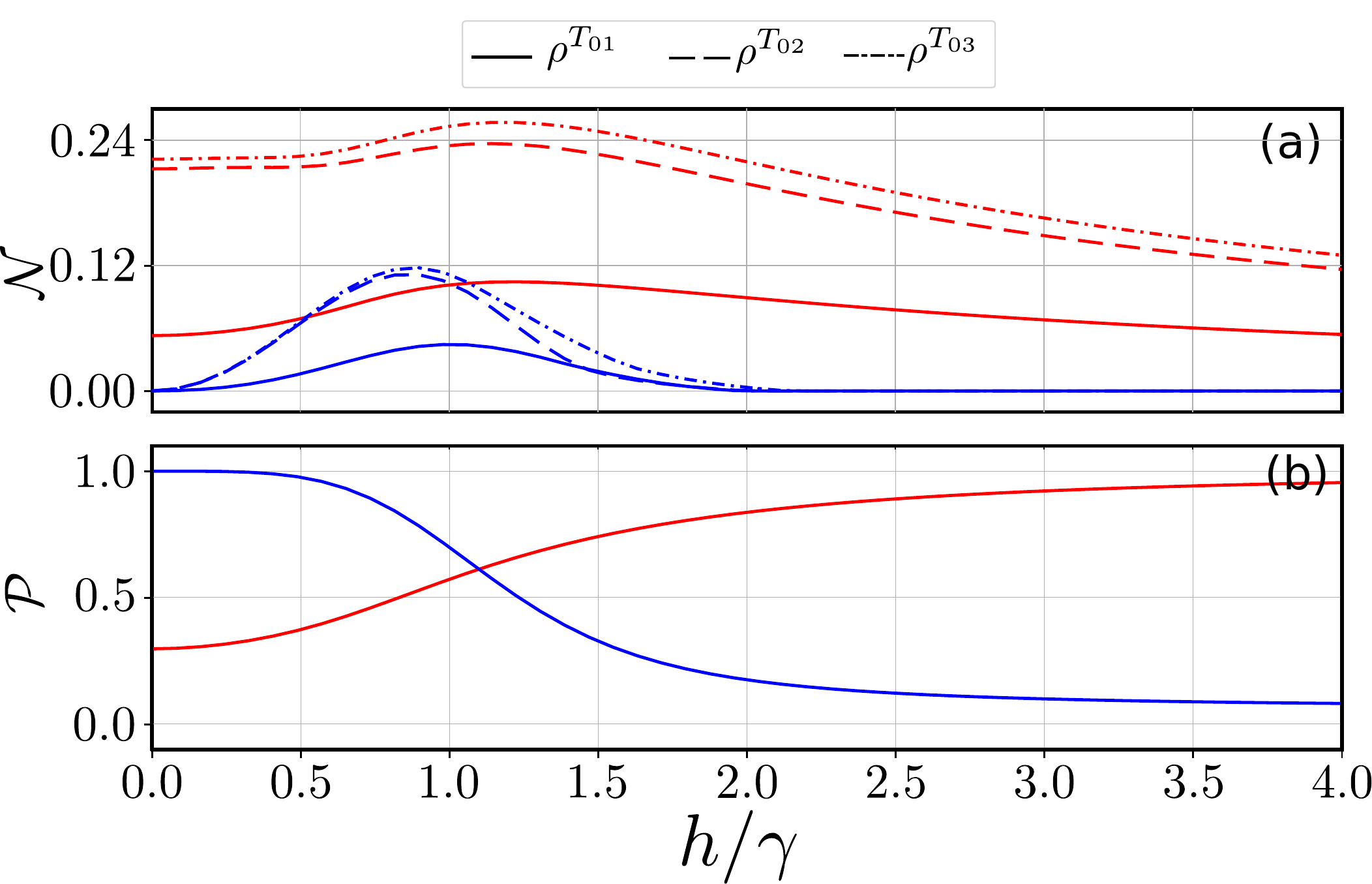}
	\caption{(a) The steady-state entanglement negativity, defined in Eq.~\eqref{Neg}  and (b) purity, $\mathcal{P}=\Tr[\rho^2]$,  for the $\sigma^x\sigma^x$-model (upper, red curves) and the $\sigma^z\sigma^z$-model (lower, blue curves). Comparing with the results of Figs.~\ref{Cost_z_fig} and \ref{Cost_x_fig} we observe a correlation between a large negativity and poor accuracy of the neural network.}
	\label{Ent}
\end{figure}

To further understand the convergence of these approaches we investigate the entanglement present in the steady-states of both models over a range of parameters. Contrary to pure states, quantifying the amount of correlations that are present in a mixed state isn't as simple as just calculating the entanglement entropy between two halves of the system~\cite{Henderson2001}. For our purposes we find that the negativity~\cite{Vidal2002, Modi2012}

\begin{equation}\label{Neg}
\mathcal{N} = \frac{||\rho^{T_A}|| - 1}{2},
\end{equation}
provides a useful measure of the correlations which are difficult to represent using the LDM approach described above.  
Here, $||\rho^{T_A}||$ denotes the trace norm of the partially transposed density matrix with the transpose taken over the degrees of freedom labeled by $A$. This quantity gives a measure for the separability of a state. If two subsystems are entangled, the partial transpose can lead to negative eigenvalues, which leads to a trace norm greater than one and hence a non-zero negativity. 

Panel (a) of Fig.~\ref{Ent} shows the negativity for both models considered in this paper for the three different possible bipartitions of a four-site system. In the case of the simpler model in Eq.~\eqref{TFI_z} with the $\sigma^z\sigma^z$ interaction we see a clear peak in the negativity at around $h\sim0.9\gamma$ for all partitions and a fast decay to zero at values of h above and below this point. This is because of the two limiting cases described in Eqs.~\eqref{eqn:pureh0} and~\eqref{eqn:mixedhinf} which both have zero negativity. The peak corresponds well to the range of $h$-values which were the the most difficult to find convergence with the neural networks.

In case of the more difficult model described in Eq.~\eqref{TFI_x} with $\sigma^x\sigma^x$ interactions we see a much higher negativity across the whole range of values of $h$. This ties in well with our experience that this model is much harder to represent using both the LDM and NDM approaches across the board, with generally slight improvements around $h=0$ and $h \gtrsim 4\gamma$. We see that there is no region where the negativity reaches zero. This model does not have a simple product state steady-state anywhere in the observed parameter range.

\section{Conclusion} \label{sec_Con}

In summary we have proposed a NNQS ansatz which compactly represents density matrices in Liouville space, allowing us to find the steady-state of lattice models described by a Markovian master equation. This LDM approach was shown to be able to calculate the steady-state of a 1D open transverse field Ising model with 6 and 16 sites.  The results were compared to the powerful NDM ansatz as implemented in NetKet. We found that our approach is always able to reach a comparable accuracy and in many cases is better able to find steady-state, especially when it contains a lot of correlations. We were able to show that the accuracy of this approach is able to be systematically improved by increasing the number of hidden units and hence free parameters in the ansatz. 

This permits a clear understanding of the class of states accessible to the LDM ansatz. As we show in Figs.~\ref{Cost_z_fig} and~\ref{Cost_16}, the most difficult regions to find convergence are very strongly correlated with parameters where the true steady-state has high negativity. This is in contrast with the NDM approach where there is difficulty representing mixed states with no correlations  leading to a plateaus in the cost function which  do not significantly decrease as the steady-state becomes more separable.

The results shown here are just a starting point for examining the usefulness of neural network approaches to finding the steady-state of open quantum systems. The RBM ansatz we used is the simplest possible architecture and extending the approach here to deep networks such as as deep RBMs~\cite{hu2016}, Recurrent Neural Networks~\cite{Sherstinsky2020}, or transformers~\cite{Vaswani2017}, which have already proven successful in closed system, provides a route to improving both the accuracy and numerical efficiency. The models studied here are also very simple and are accessible by other methods such as tensor network based techniques. However, the lack of an underlying lattice geometry for these neural networks can be exploited to study models with long range interactions and in higher dimensions which are much more difficult to simulate using other approaches.

\acknowledgments

We acknowledge useful discussions with Andrew Daley, Damian Hoffman, Daniela Pfannkuche and Filippo Vicentini. SK acknowledges financial support from EPSRC (EP/T517938/1).

\appendix
\section{Stochastic Reconfiguration} \label{app:SR}

{The Stochastic Reconfiguration update method was originally proposed in the context of attempting to develop variational quantum Monte-Carlo algorithms which avoid the sign problem~\cite{Sorella1998, Sorella2001}.}
The derivation of the update in this appendix is based on the ones found in Refs.~\cite{Becca2017,Glasser2018,Park2020}. We recommend especially~\cite{Becca2017} for various additional ways of deriving SR and Ref.~\cite{Park2020} for an in-depth discussion of the properties of the quantum Fisher matrix which plays an important role in SR.

Consider an ansatz function $\ket{\rho_{\alpha}}$ with variational parameters $\alpha$. For the derivation, we assume that the ansatz is normalized: $||\rho_{\alpha}||^2 = \langle\rho_{\alpha}|\rho_{\alpha}\rangle = 1$. 

We can use this to define a semi-orthogonal basis consisting of $\ket{\rho_{i=0,\alpha}} = \ket{\rho_{\alpha}}$, the ansatz function, as well as its derivatives

\begin{equation}
\ket{\rho_{i,\alpha}} = (O_i - \langle O_i\rangle)\ket{\rho_{0,\alpha}},
\end{equation}
where $\langle O_i\rangle = \bra{\rho_{0,\alpha}} O_i \ket{\rho_{0,\alpha}}$ and $O_i$ is the logarithmic derivative operator $O_i(s) = \frac{1}{\rho_{\alpha}(s)}\frac{\partial}{\partial_{\alpha_i}}\rho_{\alpha}(s)$. 

A variation by a small parameter shift $\gamma$ then yields:
\begin{equation}
\ket{\rho_{\alpha + \gamma}} \approx\sum^{N_p}_{i=0} \gamma_i\ket{\rho_{i,\alpha}}.
\end{equation}
Let us now suppose we are given some Liouvillian $L$, then real-time evolution  over some small time-step $\delta t$ is given by $\text{e}^{-\delta tL}\ket{\rho_{\alpha}}$. Expanding this around a small time-step yields

\begin{equation}
\text{e}^{-\delta tL}\ket{\rho_{0,\alpha}} \approx (1 - \delta tL)\ket{\rho_{0,\alpha}}.
\end{equation}
We now have:

\begin{align*}
\ket{\rho_{\alpha + \gamma}} &\approx\sum^{N_p}_{i=0} \gamma_i\ket{\rho_{i,\alpha}}\\
\ket{\tilde{\rho}_{0,\alpha}} &\approx (1 - \delta tL)\ket{\rho_{0,\alpha}}.
\end{align*}
The idea is now to project both sides into the non-orthogonal basis and ask, under which conditions they become equal \cite{Glasser2018}:

\begin{align*}
&\langle\rho_{i,\alpha}|\tilde{\rho}_{0,\alpha}\rangle = \langle\rho_{i,\alpha}|\rho_{\alpha + \gamma}\rangle\\
&\Rightarrow\langle\rho_{i,\alpha}|(1 - \delta tL)\ket{\rho_{0,\alpha}} =\langle\rho_{i,\alpha}|\sum^{N_p}_{j=0} \gamma_j\ket{\rho_{j,\alpha}}\\
&\Rightarrow  - \delta t\langle\rho_{i,\alpha}|L\ket{\rho_{0,\alpha}} =\sum^{N_p}_{j=1} \gamma_j\langle\rho_{i,\alpha}|\rho_{j,\alpha}\rangle\\
\end{align*}
The left-hand side now reads

\begin{equation}
- \delta t\langle\rho_{0,\alpha}|(O^{*}_i - \langle O^{*}_i\rangle)L\ket{\rho_{0,\alpha}} = - \delta t(\langle O^{*}_iL\rangle - \langle O^{*}_i\rangle \langle L\rangle),
\end{equation}
while the right-hand side reads

\begin{align*}
&\sum^{N_p}_{j=1} \gamma_j\langle\rho_{0,\alpha}|(O^{*}_i - \langle O^{*}_i\rangle)(O_j - \langle O_j\rangle)\rho_{0,\alpha}\rangle\\
&\sum^{N_p}_{j=1} \gamma_j (\langle O^{*}_iO_j\rangle - \langle O^{*}_i\rangle\langle O_j\rangle)
\end{align*}
Equating both sides yields:

\begin{equation}
- \delta t(\langle O^{*}_iL\rangle - \langle O^{*}_i\rangle \langle L\rangle) = \sum^{N_p}_{j=1} \gamma_j (\langle O^{*}_iO_j\rangle - \langle O^{*}_i\rangle\langle O_j\rangle).
\end{equation}
Identifying the forces $f_i=\langle O^{*}_iL\rangle - \langle O^{*}_i\rangle \langle L\rangle$ and the quantum Fisher matrix $S_{i,j}=\langle O^{*}_iO_j\rangle - \langle O^{*}_i\rangle\langle O_j\rangle$ we can write this more succinctly

\begin{equation}
-\delta t{f}=S{\gamma}.
\end{equation}
The solution of this system of linear equations ${\gamma}$ are the parameter updates which most closely resemble a step in real-time of size $\delta t$.

\section{Markov Chain Monte Carlo Sampling}
\label{MCMC}

Since the size of the required Liouville space grows exponentially fast with the system size, it quickly becomes impossible to calculate expectation values and gradients exactly. This requires us to draw sample states from the total space in a way that follows the true probability distribution parametrized by the variational ansatz.

The algorithm we employ to generate the samples is the Metropolis-Hastings algorithm. It can be used to draw samples from an unknown distribution, as long as we have a function which is at least proportional to that distribution. We achieve this by realizing that the relative probability of two samples $\sigma$, $\frac{p(\sigma_0)}{p(\sigma_1)}$, is independent of the exact normalization of the distribution.

Each sample is saved and generated iteratively in such a way, that the next sample depends only on the current sample. In this way a Markov Chain of samples is created that can be used to estimate expectations and gradients.

The full algorithm is as follows.
Let $\rho_\alpha(x)$ be the function that is proportional to the desired distribution $|\rho|^2$. Further, let $Q(x|y) = Q(y|x)$ be the symmetric, conditional probability distribution that determines the jumps. Note, that we never explicitly calculate $Q(x|y)$. In case of flipping one random spin, the probability to flip any spin is always $1/N$ independent of the current configuration. Hence, $Q(x|y) = Q(y|x)$ is trivially fulfilled. 

\begin{enumerate}
\item Choose a random spin configuration $\sigma_0$. 

\item Use $Q(x|y)$ to suggest a new configuration $\sigma_1$.

\item Calculate the relative probability $p = \left|\frac{\rho_\alpha(\sigma_1)}{\rho_\alpha(\sigma_0)}\right|^2$

\item Generate a random number $r$ between 0 and 1 and compare with the relative probability. If $r<p$, accept the new configuration and add it to the list. Otherwise add the old sample to the list again.

\item Repeat the process until a sufficient number of samples has been generated

\item Calculate the average of the operators and gradients from this set as required
\end{enumerate}

There are a few pitfalls to look out for. {Firstly}, since each sample is generated as a small change from the previous sample, neighboring samples are strongly correlated and the full chain may not faithfully represent the true distribution. {Secondly}, the first sample of a chain is randomly chosen from all possible states with a uniform probability, while it might have an infinitesimally small weight in the true distribution. Hence, the first sample and a number of samples following it, can drastically skew the distribution represented by the chain. {Thirdly}, it is not at all obvious, when enough samples have been generated to represent the true distribution.

These problems have been known for a long time and have been discussed in the literature~\cite{Gelman1992,Katzgraber2011,Becca2017}. The first problem is best addressed by only keeping one in every $n$ of the samples. This ensures that each sample in the resulting chain is independent of the previous one, see~\cite{Katzgraber2011,Becca2017}. The second problem can be addressed in a similar fashion. Instead of taking a randomly selected sample, one generates an entire chain again and uses the last entry as the starting point for the final Markov Chain. This ensures that the starting point of the simulation lies in a corner of the state space with finite weight. The third problem is addressed in e.g.~Refs.~\cite{Gelman1992,Gelman1997,Vats2020}. They proposed to estimate the variance between several Markov Chains and within these Chains, arguing that their ratio should approach unity in the limit of perfect convergence or infinite samples. This gives a good estimate of how close the samples are to representing the true probability distribution. This measure is known in the literature as the Gelman-Rubin R value.

\end{document}